\documentstyle[12pt]{article}

\newcommand{\mincir}{\raise -2.truept\hbox{\rlap{\hbox{$\sim$}}\raise5.truept
\hbox{$<$}\ }}
\newcommand{\magcir}{\raise -2.truept\hbox{\rlap{\hbox{$\sim$}}\raise5.truept
\hbox{$>$}\ }}
\newcommand{\minmag}{\raise-2.truept\hbox{\rlap{\hbox{$<$}}\raise 6.truept\hbox
{$>$}\ }}
\newcommand{\be}{\begin{equation}}
\newcommand{\ee}{\end{equation}}
\newcommand{\ba}{\begin{eqnarray}}
\newcommand{\ea}{\end{eqnarray}}
\newcommand{\brr}{\begin{array}}
\newcommand{\err}{\end{array}}
\newcommand{\bc}{\begin{center}}
\newcommand{\ec}{\end{center}}

\newcommand{\hm}{\,h^{-1}{\rm Mpc}}

\begin{document}
\baselineskip 0.7cm

\begin{titlepage}

  \begin{flushright}
    TUM-HEP-242/96
    \\
    SFB-375/95
    \\
   April 1996
  \end{flushright}

\begin{center}
  {\Large \bf Light Gravitinos as Mixed Dark Matter}
\vskip 0.5in
{\large
    Stefano Borgani,$^{a,b)}$  Antonio Masiero$^{c)}$ 
and Masahiro Yamaguchi$^{d)}$\footnote{
On leave of absence from Department of Physics, Tohoku University, Sendai
980-77, Japan}
}
\vskip 0.4cm 
{\it a) INFN, Sezione di Perugia \\
    Dipartimento di Fisica, Universit\`{a} di Perugia \\
    via A. Pascoli, I-06100 Perugia, Italy} \\

{\it b) SISSA-International School for Advanced Studies \\
       via Beirut 2-4, I-34013 Trieste, Italy} \\

{\it c) Dipartimento di Fisica,  Universit\`{a} di Perugia and  \\
   INFN, Sezione di Perugia, via A. Pascoli, I-06100 Perugia, Italy} \\

{\it d) Institute f\"{u}r Theoretische Physik \\
Physik Department, Technische Universit\"{a}t M\"{u}nchen \\
D-85747 Garching, Germany}

\vskip 0.5in

\abstract  In theories with a gauge-mediated mechanism of supersymmetry 
breaking the gravitino is likely to be the lightest superparticle and,
hence, a candidate for dark matter. We show that the decay of the 
next-to-lightest superparticle into a gravitino can yield a non-thermal 
population of gravitinos which behave as a hot dark matter component. 
Together with the warm component, which is provided by the population 
of gravitinos of thermal origin, they can give rise to viable schemes of 
mixed dark matter. This realization has some specific and testable
 features both in particle physics and astrophysics. We outline under 
which conditions the mechanism remains viable even when R parity is broken.

\end{center}
\end{titlepage}

A deeper comprehension of the mechanism (origin, nature and scale) giving rise
to the breaking of local supersymmetry (SUSY) still represents the major
challenge we are facing on our way to construct realistic models of low-energy
N=1 supergravity models \cite{nilles}.  So far, the most conventional approach
makes use of a so-called ``hidden" sector \cite{nilles} which is made
responsible for SUSY breaking at a large scale ($10^{10}-10^{11}$ GeV) with a
gravitino mass of $O(10^2-10^3$ GeV) and the gravitational interactions 
representing the messenger of the SUSY breaking from the hidden to the 
observable sector. 
  The alternative view that SUSY may be
broken in a ``secluded" sector at a much lower scale with gauge instead 
of gravitational forces responsible for conveying the breaking of SUSY 
to the observable sector
 had already been critically
considered in the old days of the early constructions of SUSY models and has
raised a renewed interest recently with the proposal by 
Refs.~\cite{dineetal,dine-old,dvalietal} where some guidelines for
 the realization
 of low-energy SUSY
breaking are provided.  In these schemes, the gravitino mass ($m_{3/2}$) loses
its role of fixing the typical size of soft breaking terms and we expect it to
be much smaller than what we have in models with a hidden sector.  Indeed,
given the well-known relation \cite{nilles} between $m_{3/2}$ and the scale of
SUSY breaking $\sqrt{F}$, i.e.\ $m_{3/2}=O(F/M)$, where $M$ is the
reduced Planck scale, we expect $m_{3/2}$ in the keV range for a scale
$\sqrt{F}$ of $O(10^6$ GeV) that has been proposed in models with low-energy
SUSY breaking in a visible sector.

In this letter we study some implications of SUSY models with a light
gravitino (in the keV range) in relation with the dark matter (DM)
problem. That such a light gravitino is
very likely to be the lightest superparticle (LSP) and, hence, at least
in models with $R$ parity conservation, a good candidate
for DM has already been known for quite some time now (since the early 
analysis of Pagels and Primack \cite{PagelsPrimack}). The new point of 
our analysis is that there actually exist two different populations of
 gravitinos
which are relic of the early Universe. First, we have gravitinos which
through their 1/2 helicity component were in thermal equilibrium at some
early epoch and, as we mentioned, are known  
\cite{PagelsPrimack} to be an interesting warm DM (WDM) candidate.  Then, we
have a kind of ``secondary population" of gravitinos, which result from
the decay of the next-to-the-lightest superparticle (NSP), presumably the
lightest neutralino.  They have a non-thermal distribution and exhibit
features for the structure formation which are similar to those of a
standard hot light neutrino in the tens of eV range. We will try to
clarify in this letter under what conditions these two populations of
gravitinos can give rise to a tenable scheme of mixed DM (MDM).  It will
turn out that viable MDM realizations within the frame with light
gravitinos that we envisage here lead to characteristic features both in
the cosmological and particle physics contexts, making these models
testable against astrophysical observations and future accelerator
experiments.  In particular, on the astrophysical side, we find a
relatively large $^4$He abundance (corresponding to slightly more than
three neutrino species), a suppression of high redshift galaxy formation
with respect to the cold dark matter (CDM) scenario and a free-streaming
scale of the non-thermal (``secondary") gravitinos independent of
$m_{3/2}$, but sensitive to the NSP mass (with important consequences on
the large scale structure formation).  As for the particle physics
implications, the implementation of a MDM scheme imposes severe
constraints on the SUSY particle spectrum.  For instance the lightest
neutralino (NSP) should be an essentially pure gaugino and sfermions have
to be rather heavy (in the TeV range). More specific features will be
discussed below.

As we previously mentioned, a stable particle with a mass in the keV
range, like the light gravitino we are discussing here, was considered
already long time ago as a possible warm DM candidate
\cite{PagelsPrimack},  i.e.\ a variant of hot DM but becoming
non-relativistic at a much earlier epoch and, hence, having a much
smaller free-streaming scale of $O(1$ Mpc). Cosmological scenarios based
on WDM were considered during the early '80s \cite{wdm}, soon after the
shortcomings of HDM for large--scale structure formation were recognized.
However, a purely WDM scenario seems to share the main drawbacks of
standard CDM, once normalized on large scales to match the amplitude of
the cosmic microwave background anistotropies \cite{cmb}: {\em (a)} a
spectrum of density fluctuations which is too steep on $\sim
20\hm$ \footnote{$h$ is the Hubble parameter in units of $100\,{\rm
km\,s^{-1}Mpc^{-1}}$; $0.5\mincir h\mincir 1$ from observations; $h=0.5$
is usually taken when considering critical density cosmological models
with $\Omega_0=1$, in order not to conflict with constraints on the age
of the Universe.} scales with respect to that  measured for the
distribution of galaxies and galaxy clusters \cite{pkobs}, and {\em (b)} too
large fluctuations on  $\sim 10\hm$ scales, resulting in an
overproduction of galaxy clusters \cite{clth}. Colombi et al. \cite{colo} have
recently considered cosmological scenarios based on
WDM. As a main conclusion,  they found that a viable WDM
candidate should have a mass--to--temperature ratio, $m_x/T_x$, twice
that of the light neutrinos required by the HDM model. For larger and
smaller values of this ratio the WDM model rapidly falls into the CDM and
HDM cases, respectively, thus requiring some degree of tuning for it to
be a substantial improvement with respect of CDM. 

The difficulty for pure CDM frames has paved the way to a
revival of the MDM scenarios \cite{MDM} where the cold and hot DM coexist, with
a certain ratio of composites, in most cases of which the latter is supposed to
be one (or more than one) light massive neutrino. In this scenario the
presence of light neutrinos suppresses the growth of density
fluctuations in the cold component on scales smaller than their
free--streaming length. This goes in the right direction of generating a
shallower spectrum, while keeping the fluctuation amplitude on the
cluster mass scale to a more adequate level.

However, it should be kept in mind that the cold+hot DM (CHDM) scenario
is only one of the possibilities to implement the MDM idea. Another
option that has been thoroughly investigated recently \cite{volatile} is
that the hot thermal component is replaced by a volatile component made
of particles with high {\it rms} velocity, which derive from the decay of
a heavier particle.  Such models based on cold+volatile DM (CVDM) differ
from the more conventional CHDM schemes in that they involve a component
which has a non-thermal phase space distribution function.  An
interesting implementation of the CVDM proposal is found in schemes where
one and the same particle can play the twofold role of cold (or warm) and
volatile component.  An example was considered in Ref.
\cite{axino-volatile}, where the fermionic partner of the axion, the
axino, contributed the LSP.  The decoupling of the axinos from the
thermal bath yields the cold (or, rather, warm) component of the DM,
whilst the axinos coming from the NSP decay constitute the volatile
component which plays a role similar to that of light hot neutrinos as
far as large scale structure formation is concerned.

Here we point out that, thanks to a mechanism similar to that which was
exploited in the abovementioned axino example, the light gravitinos can account
for the warm as well as the volatile component, provided the NSP abundance
before decay is large enough to yield a significant amount of non-thermal
gravitinos. 

The helicity 3/2 component of gravitino has couplings of gravitational
 strength.  If the Universe underwent the inflationary era,
which we assume hereafter, its abundance was completely diluted during
the inflation and is never produced later, so as to constitute a
significant portion of the mass density of the Universe.  Thus the
helicity 3/2 component plays no role in cosmology in this light gravitino
case.  On the other hand, the helicity 1/2 component, or the longitudinal
mode of the gravitino, has much stronger interaction \cite{Fayet1} 
when the gravitino
is light having  SUSY broken at a low energy scale.  This is
because it is essentially a Goldstino associated with the SUSY breaking,
whose decay constant is given by the SUSY breaking scale $\sqrt{F} \sim
(m_{3/2} M)^{1/2}$.  Indeed the explicit form of the interaction is
\cite{Fayet2} 
\begin{eqnarray}
  {\cal L}_{eff}= \frac{m_{\lambda}}{8 \sqrt{6} m_{3/2} M}
                  \bar \psi [\gamma_{\mu},\gamma_{\nu}] \lambda F_{\mu \nu}
        + \frac{m_{\chi}^2-m_{\phi}^2}{\sqrt{3}m_{3/2}M} \bar \psi \chi_L
             \phi^* +h.c.,
\end{eqnarray}
where $\psi$ represents the helicity 1/2 component gravitino (the Goldstino) 
and
$m_{\lambda}$, $m_{\chi}$ and $m_{\phi}$ are the masses of the gaugino, the
chiral fermion and its superpartner, respectively.

The helicity 1/2 component of the gravitino (hereafter we call it simply the
gravitino) with such much stronger interaction than the
gravitational one can be in thermal equilibrium at an early epoch of the
Universe.  To see this, let us consider production/destruction of gravitinos
a) by scattering ($a +b \leftrightarrow c +\psi$), and b) by decay and
inverse-decay of a superparticle into a gravitino ($a \leftrightarrow b
+\psi$).  The total cross section for $a+b \rightarrow c + \psi$ was calculated
in Ref.~\cite{MMY}, being roughly
\begin{eqnarray}
      \Sigma_{tot} \simeq \frac{1}{\pi} \frac{m_{\tilde g}^2}{m_{3/2}^2 M^2},
\end{eqnarray}
where $m_{\tilde g}$ is the gluino mass.  The resulting interaction rate is
\begin{eqnarray}
\Gamma_{scatt} \simeq \Sigma_{tot} n_{rad}
\end{eqnarray}
with $n_{rad}=\zeta(3)/\pi^2 T^3$ being the number density for one massless
degree of freedom.  Comparing this with the expansion rate of the Universe $H$,
one finds that the interaction is effective to keep the equilibrium
($\Gamma_{scatt} \magcir H$) until the temperature becomes\footnote{For $T
  \mincir m_{\tilde g}$ the above estimate for $\Sigma_{tot}$ will not be
  accurate. However, it will not change our argument drastically.}
\begin{eqnarray}
   T \simeq O(10^2) \mbox{GeV} 
     \left( \frac{m_{3/2}^2}{1\mbox{keV}} \right)
     \left( \frac{1\mbox{TeV}}{m_{\tilde g}} \right)^2.
\end{eqnarray}
Next consider the decay and inverse-decay process.  
The decay width of a superparticle (R-odd particle) $ a$  into its
superpartner $b$ and a gravitino is given by \cite{GHH,MMY}
\begin{eqnarray}
   \Gamma (a \rightarrow b \psi)
 =\frac{1}{48 \pi}\frac{m_{ a}^5}{m_{3/2}^2 M^2}, 
\label{width}
\end{eqnarray}
where $m_{a}$ is the mass of $a$.
Suppose that, for simplicity,  all
superparticles in the MSSM have the same mass, $m_{S}$.  Given a temperature
$T>m_{S}$,  the interaction rate for this process is estimated as
\begin{eqnarray}
 \Gamma_{decay}&=& \langle \Gamma \rangle \times \frac{n_S}{n_{\psi}} 
\nonumber \\
               &\simeq& g_S \frac{m_S}{T} \Gamma( a \rightarrow b \psi)  
\nonumber \\
               &\simeq& \frac{g_S}{48 \pi} \frac{m_S^6}{m_{3/2}^2 M^2 T}.
\end{eqnarray}
Here $g_S$ represents the effective degrees of freedom of the superparticles in
the MSSM, $\sim 100$, and $n_S$ is the number density of the 
superparticles.  From this it follows that
\begin{eqnarray}
\frac{\Gamma_{decay}}{H} \simeq \frac{g_S}{50 g_*(T)^{1/2}} 
                              \frac{m_S^6}{m_{3/2}^2 M T^3},
\end{eqnarray}
where $g_{*}(T)$ represents the effective massless degrees of freedom.
Thus one may conclude that for $m_S \magcir O(10^2) \mbox{GeV}
(m_{3/2}/1\mbox{keV})^{3/2}$, the decay and inverse-decay process is
effective to maintain the equilibrium of the gravitinos as long as the
superparticles are relativistic.  As they become non-relativistic, the
interaction rate drops exponentially, $\sim e^{-m_S/T}$, due to the
Boltzmann suppression of the number density of the superparticles.  In
order to determine the freeze-out temperature $T_f$ precisely, one must
explicitly take into account  the superparticle mass spectrum when
integrating numerically the Boltzmann equation. Here we will not attempt
to do so, rather we just estimate that the effective massless degrees of
freedom at $T_f$ is $g_*(T_f) \simeq 100 -200$, admitting a factor 2
ambiguity.  This is enough for the purpose of this paper.

Suppose that the gravitinos were once in thermal equilibrium and were
frozen out with $g_*(T_f)$ given above.  We assume that the gravitino is the
LSP and stable, which is guaranteed by $R$ parity conservation.  We will
discuss later the case where $R$ parity is not conserved.  Following the
standard procedure, the density parameter $\Omega_{th}$ contributed by relic 
thermal gravitinos is
\begin{eqnarray}
   \Omega_{th}h^2 =1.17 \left( \frac{m_{3/2}}{1\mbox{keV}} \right) 
               \left( \frac{g_{*}(T_f)}{100} \right)^{-1} ,
\end{eqnarray}
or
\begin{eqnarray}
      m_{3/2}=0.85 \mbox{keV} 
         \left( \Omega_{th} h^2 \right)
         \left( \frac{g_*(T_f)}{100} \right).
\end{eqnarray}
Therefore, a
gravitino in the abovementioned keV range provides a significant portion
of the mass density of the present Universe and behaves as a
WDM candidate, as was mentioned previously.

We now turn to another contribution of the gravitino abundances, namely
gravitinos produced from massive particle decays after the freeze-out.  To be
specific, we consider the gravitinos produced by the non-thermal decays of 
binos, the superpartner of the $U(1)_Y$ gauge boson,  assuming that it is
the NSP, i.e.\ the lightest among the superparticles in the MSSM sector.

>From Eq.~(\ref{width}), one finds that the decay width of a bino into a photon 
and a gravitino is
\begin{eqnarray}
   \Gamma (\tilde b \rightarrow \gamma \psi)
 &=& \cos^2 \theta_W \times \frac{1}{48 \pi} 
    \frac{m_{\tilde b}^5 }{m_{3/2}^2 M^2} 
\nonumber \\
 &=& 2.15 \times 10^{-20} \mbox{GeV} 
    \left( \frac{m_{\tilde b}}{30 \mbox{GeV}} \right)^5
    \left( \frac{1 \mbox{keV}}{m_{3/2}} \right)^2, \label{bino-width}
\end{eqnarray}
where $m_{\tilde b}$ is the bino mass and $\theta _W$ stands for the weak mixing
angle.

Equating the above decay rate with the expansion rate of the Universe, 
one determines the bino decay temperature $T _D$:
\begin{eqnarray}
   T_D=0.280 \mbox{GeV} \times g_*(T_D)^{-1/4} 
       \left(\frac{m_{\tilde b}}{30 \mbox{GeV}} \right)^{5/2}
       \left(\frac{1 \mbox{keV}}{m_{3/2}}\right).
\end{eqnarray}
Hence the bino decay takes place well before nucleosynthesis starts.

Each bino produces one gravitino at decay. From this, it follows that the
abundance of the non-thermal gravitinos coming from the bino decays is
\begin{eqnarray}
\Omega_{non\mbox{-}th}=\frac{m_{3/2}}{m_{\tilde b}} \Omega_{\tilde b}
\end{eqnarray}
where $\Omega_{\tilde b}$ represents the would-be bino mass density at present,
if it were stable.  For example, for $m_{3/2}=0.5$ keV, $m_{\tilde b}=30$ GeV,
one needs $\Omega_{\tilde b}=1.2 \times 10^7$ to achieve
$\Omega_{non\mbox{-}th}=0.2$.

This large abundance required for the next-to-the-lightest superparticle (NSP)
imposes severe constraints on mass spectrum of superparticles.  First of all,
the NSP must be almost pure gaugino: a non-negligible contamination of higgsino
component would allow it to annihilate through, for example, $s$-channel $Z$
boson exchange so that one would not expect such a large $\Omega_{\tilde b}$.
Second, it requires that sfermions (i.e.\ squarks and sleptons) must be
very heavy.  The relic abundance of the bino is estimated as
\cite{OliveSrednicki} 
\begin{eqnarray}
  \Omega_{\tilde b} h^2 \simeq 1 \times 10^{-6} 
            \frac{(m_{\tilde f}^2 +m_{\tilde b}^2)^2}{m_{\tilde b}^2}
            \mbox{GeV}^{-2},
\end{eqnarray}
where we have assumed for simplicity that all sfermions have the same mass
$m_{\tilde f}$. For $m_{\tilde b}=30$ GeV and $h=0.5$, the sfermion mass
$m_{\tilde f}$ must lie around 7 TeV to obtain the
abovementioned value $\Omega_{\tilde b}=1.2 \times 10^7$. 

This mass pattern provides a non trivial constraint for a model of 
low-energy SUSY breaking.  In
the model of Ref.~\cite{dineetal}, the messenger sector connects the visible
sector with the supercolor sector in such a way that gauginos can acquire their
masses at one-loop level whereas squarks and sleptons get mass squared at
two-loop level. The ratio between the gaugino and sfermion masses is 
model-dependent. Indeed, it is strictly related to the specific pattern of the
mass matrix of the messenger fields.  One can envisage solutions 
where a mass pattern suitable for the scheme that we propose here is 
actually implemented.\footnote{We thank G. Giudice for discussions on 
this point.}   

Thus far, we have obtained two different populations of the gravitino DM. One
is the thermal gravitino which gives the WDM, and the other is the
non-thermal gravitino whose properties are similar to the HDM.  To clarify
the latter point, we calculate the redshift $Z_{nr}$ at which the non-thermal
gravitino becomes non-relativistic.  The momentum of the non-thermal gravitino 
gets red-shifted as
\begin{equation}
   k_{phys}(t)=\frac{m_{\tilde b}}{2} \frac{a_D}{a(t)}  \label{red-shift}
\end{equation}
where $a(t)$ is the 
expansion
factor at a given time $t$ and $a_D$ is its
value at the bino decay.  Let $a_{nr}$ be the expansion factor when the
gravitino becomes non-relativistic, i.e. when $k_{phys}=m_{3/2}$. From 
Eq.~(\ref{red-shift}), it follows that  
\begin{equation}
   m_{3/2} =\frac{m_{\tilde b}}{2} \frac{a_D}{a_{nr}}.
\end{equation}
Therefore, it turns out that 
\begin{eqnarray}
 Z_{nr} +1 &\equiv& \frac{a_0}{a_{nr}}
       = \frac{a_0}{a_D} \frac{a_D}{a_{nr}} 
\nonumber \\
     &=& \left( \frac{g_{*S}(T_D)}{g_{*S}(T_0)} \right)^{1/3}
         \frac{T_D}{T_0} \frac{2 m_{3/2}}{m_{\tilde b}} 
\nonumber \\
    &= &6.42 \times 10^4 
             \left(\frac{ g_{*}(T_D)}{20} \right)^{1/12} 
             \left(\frac{m_{\tilde b}}{30 \mbox{GeV}} \right)^{3/2}.
\label{Znr}
\end{eqnarray}
For a heavier $m_{\tilde b}$, the temperature at its decay $T_D$ will be larger
than the freeze-out temperature of the bino, $\sim m_{\tilde b}/20$ and then
the above argument for $Z_{nr}$ should be modified. We will not detail this
here, since a heavier $m_{\tilde b}$ yields a larger $Z_{nr}$, which is
presumably less attractive from the viewpoint of large scale structure
formation.

It is interesting to compare this with the redshift at which the
thermal gravitinos get non relativistic. Since $Z$ is related to the gravitino
temperature $T_{3/2}$ according to
\begin{eqnarray}
       Z+1=\frac{a_0}{a}
   = \left( \frac{g_*(T_f)}{g_{*S}(T_0)} \right)^{1/3}
     \frac{T_{3/2}}{T_0},
\end{eqnarray}
and the thermal gravitinos become  non-relativistic around when their
temperature becomes $m_{3/2}/3$, it turns out that 
\begin{eqnarray}
       Z_{nr} 
  & \sim& \left( \frac{g_*(T_f)}{g_{*S}(T_0)} \right)^{1/3}
     \frac{m_{3/2}/3}{T_0} 
\nonumber \\
  & =& 4.14 \times 10^6 \times  
     \left( \frac{g_*(T_f)}{100} \right)^{1/3}
     \left( \frac{m_{3/2}}{1\mbox{keV}} \right). 
\end{eqnarray}

 Hence  thermal gravitinos become non-relativistic much earlier than
the non-thermal ones.  

Once $Z_{nr}$ is known, one can estimate  the free streaming length until the
epoch of the matter-radiation equality, $\lambda_{FS}$, which represents
a quantity of crucial relevance for the formation of large--scale cosmic
structures. If $v(t)$ is the typical velocity of a DM particle at the
time $t$, then
\begin{eqnarray}
     \lambda_{FS}&\equiv& \int_0^{t_{eq}} \frac{v(t)}{a(t)} dt 
\nonumber \\
                 &=& 2 t_0 \times \frac{Z_{eq}^{1/2}}{Z_{nr}}
                         [1+\ln (Z_{nr}/Z_{eq})] 
\nonumber \\
                 &=& 6.08 \times 10^5 \times Z_{nr}^{-1} \mbox{Mpc}
                  [1+\ln(Z_{nr}/2.32 h^2 \times 10^4)]
\end{eqnarray}
According to this estimate, for the non-thermal gravitino with
$Z_{nr}=6.42 \times 10^4$, the free-streaming length is $\lambda \sim 30$
Mpc, which corresponds to a supercluster size.  Thus the non-thermal
gravitino in our scenario will exhibit properties similar to those of a 
regular HDM candidate, like a light thermal neutrino, as far as large 
scale structure formation is concerned. 
 On the other hand, the free-streaming
length for the thermal gravitinos is about 1Mpc (for $Z_{nr}\sim 4 \times
10^6$), which in turn corresponds to $\sim10^{12}M_\odot$.  

Therefore, this scenario corresponds to a MDM model, with warm and
volatile DM components provided by thermal and non--thermal gravitinos,
respectively. A similar scheme has been recently considered by Malaney et
al. \cite{whdm}, where the hot component is represented by ordinary light
thermal neutrinos, while the warm part is provided by sterile neutrinos.
These authors conclude from their analysis that this model is virtually
indistinguishable from the cold+hot DM scenario as far as the
large--scale structure formation is concerned. 
A crucial difference between this scheme and the one we are proposing here
lies in the fact that for the thermal neutrinos the value of $Z_{nr}$ is
fixed by their mass and, therefore, by the contributed density parameter 
$\Omega_\nu$. On the contrary, $Z_{nr}$ for the non--thermal
gravitinos does not have a one to one correspondence with $\Omega_{non-th}$
[cf. Eq.~(\ref{Znr})]. 

As for the warm component, taking $m_{3/2}\sim 200$ eV and being
\be
T_{3/2}\,\simeq \,\left({10.75\over g_*(T_D)}\right)^{1/3} T_\nu
\ee
the relation between the neutrino and the gravitino temperature, it turns out
that the ratio $m_{3/2}/T_{3/2}$ is at least 10 times larger than $m_\nu
/T_\nu$, thus showing that the warm component behaves like CDM, at least
on scales $\magcir 1\hm$. 

Pierpaoli et al. \cite{volatile} have recently considered a MDM scheme,
in which thermal axinos play the role of CDM, while non--thermal axinos
provide the volatile part. Therefore, as far as the cosmological
implications are concerned, this model behaves like the one based on
gravitinos, at least on mass scales larger than that of a galaxy. As a
main result of their analysis,  Pierpaoli et al. \cite{volatile} pointed
out that the simultaneous request of reproducing the observed abundance
of galaxy clusters \cite{clth} and of high--redshift ($Z\simeq 4$) damped
Ly--$\alpha$ systems (DLAS) \cite{dlas} requires  $\Omega_{non\mbox{-}th}\simeq
0.2$ and $Z_{nr} \mincir 10^4$. As for the cluster abundance, since it is
determined by the fluctuation amplitude on scales $\sim 10\hm$, we expect
similar predictions when replacing the cold component with the warm one.
However, since DLAS are ought to be associated with protogalaxies, the
constraint they provide refers to scales $\mincir 1\hm$. In this regime,
the effect of free--streaming in the warm component suppresses the
fluctuation growth, therefore decreasing the predicted abundance of
high--redshift DLAS. This is potentially a critical test for our DM
model, the DLAS abundance being already recognized as a non--trivial
constraint for usual cold+hot DM models \cite{dlasth}. It is however
clear that more quantitative  conclusions would at least require the
explicit computation of the fluctuation power spectrum for our
warm+volatile DM model, which is beyond the scope of this letter.
Furthermore, before definitely assessing the confidence level at which
such observational constraints rule out a model,  a further clarification
of both the reliability of available data and of the corresponding
interpretative framework are required.

As already pointed out, the value of $Z_{nr}$ for non--thermal gravitinos
in principle does not depend on $\Omega_{non\mbox{-}th}$. On the other hand, a
large amount of exotic relativistic particles around the era of the
big-bang nucleosynthesis would contribute to the energy density of the
Universe at that epoch, so as to accelerate the expansion of the Universe
and result in a significant increase of the $^4$He abundance. In our
case, both thermal and non-thermal gravitinos will contribute. It is
convenient to express such contributions in terms of the effective number
of extra generations of neutrinos defined by $\Delta N_{\nu} \equiv
\Delta \rho/\rho_{\nu}$, where $\Delta \rho$ is a contribution to the
energy density by an exotic particle and $\rho_{\nu}$ is the energy
density of a neutrino (one species). The contribution from the thermal
gravitino is easily obtained as %
\begin{equation}
  \Delta N_{\nu}^{th}
  =\left( \frac{T_{3/2}}{T} \right)^4
  =\left( \frac{g_{*}(T)}{g_{*}(T_f)} \right)^{4/3}
  =0.020 \left( \frac{200}{g_{*}(T_f)} \right)^{4/3}.
\label{DeltaNth}
\end{equation}
On the other hand, to evaluate the contribution from the non-thermal gravitino,
one should note that the energy density of the gravitino $\rho_{non\mbox{-}th}$
evolves differently from that of the neutrino $\rho_{\nu}$ only after the
gravitino becomes non-relativistic:
\begin{eqnarray}
   \Delta N_{\nu}^{non\mbox{-}th}
 = \frac{\rho_{non\mbox{-}th}}{\rho_{\nu}}\bigg |_{1 {\rm MeV}} 
 =\frac{\rho_{non\mbox{-}th}(t_0)}{\rho_{\nu}(t_0)} \frac{a_{nr}}{a_0} .
\end{eqnarray}
>From this it follows that 
\begin{eqnarray}
  \Delta N_{\nu}^{non\mbox{-}th}=0.134 \times 
         \left( \frac{\Omega_{non\mbox{-}th} h^2 }{0.05} \right)
         \left( \frac{6.42 \times 10^4}{Z_{nr}} \right).
\label{DeltaNnonth}
\end{eqnarray}
The total contribution of the gravitinos to the energy density during the
nucleosynthesis is the sum of Eqs.~(\ref{DeltaNth}) and
(\ref{DeltaNnonth}). Note that it is comparable to, or even larger than
the $2 \sigma$ upper bound for $\Delta N_{\nu} =0.16 $ coming from the
observations of the $^4$He abundances obtained recently by Olive and
Scully \cite{os}. As was cautioned by these authors, the systematic error
they used to obtain this number might be somewhat underestimated.
On the other hand, Kernan and Sarkar \cite{ks} claimed that up to $\Delta
N_\nu \simeq 1.5$ is allowed by observations of high D abundance in
high-redshift Ly$\alpha$ clouds, while no general agreement exists about
deuterium observations in high-redshift QSO's (see, e.g., ref.\cite{hsbl}
and references therein).
Keeping this in mind, we believe that it is premature to exclude our
scenario by this constraint of the $^4$He abundance.  Rather we should
emphasize that our scenario requires a significant excess of the
primordial $^4$He 
and D
abundance which might be excluded or confirmed in future.

So far, we have assumed that $R$ parity is strictly conserved.  As we will
see soon,  the light gravitino is  long lived enough to account for DM
even when $R$ parity is broken.  This contrasts with the case where
the LSP is a neutralino: unless $R$ is violated by an extremely tiny 
amount, one has to demand $R$ conservation to ensure that the LSP is a 
viable DM candidate.

 $R$ parity breaking inevitably leads to baryon or lepton number
non-conservation. Given the lightness of the gravitino that we consider 
here, it cannot decay into baryons. Hence it is stable if $R$ is 
violated in the baryonic sector.  
  Next we will estimate the lifetime of
the gravitino when the lepton number is not conserved. To be specific, we
concentrate on the case that a term $\lambda_{ijk} L_iL_j E^c_k$ with a Yukawa
coupling $\lambda_{ijk}$ appears in the superpotential as an $R$ violating
interaction, where $L$ and $E^c$ are chiral multiplets of $SU(2)_L$ doublet and
singlet leptons, respectively. The Latin subscripts represent generations and
no summation over them is taken.  At the one-loop, the gravitino decays to a
photon and a neutrino (or an anti-neutrino) through lepton-slepton loops, with
the decay rate
\begin{eqnarray}
    \Gamma (\psi \rightarrow \nu (\bar \nu) \gamma)
   &=&\frac{\alpha \alpha_{\lambda}}{96 \pi^3} \frac{ m_{3/2}}{M^2}
   \left[ m_{l_k}^2 \left( \ln (m_{\tilde l_{Lj}}/m_{l_k}) \right)^2
        +m_{l_k}^2 \left( \ln (m_{\tilde l_{Li}}/m_{l_k}) \right)^2 \right.
\nonumber \\
    &   +& \left. m_{l_j}^2 \left( \ln (m_{\tilde l_{Rk}}/m_{l_j}) \right)^2
        +m_{l_i}^2 \left( \ln (m_{\tilde l_{Rk}}/m_{l_i}) \right)^2 
 \right],
\end{eqnarray}
where $\alpha$ is the fine-structure constant,
$\alpha_{\lambda}=\lambda_{ijk}^2/4\pi$, $m_l$ the mass of the charged lepton,
and $m_{\tilde l_L}$ ($m_{\tilde l_R}$) is the mass of the left-handed
(right-handed) slepton.  The decay rate will be maximized when $\lambda_{i33}
\neq 0$ ($i=1$, or 2).  If we assume for simplicity that all charged sleptons
have the same mass 
$m_{\tilde l_L}=m_{\tilde l_R} \equiv m_{\tilde l}$, then the
decay rate reads
\begin{eqnarray}
\Gamma (\psi \rightarrow \nu (\bar \nu) \gamma)
   =\frac{\alpha \alpha_{\lambda}}{32 \pi^3} \frac{ m_{3/2}m_{\tau}^2}{M^2}
    \left( \ln (m_{\tilde l}/m_{\tau}) \right)^2
\end{eqnarray}
For $m_{\tilde l}=7$ TeV, one finds the lifetime 
\begin{eqnarray}
    \tau(\psi \rightarrow \nu (\bar \nu) \gamma)
   = 2 \times 10^{21} \alpha_{\lambda}^{-1}
     \left( \frac{m_{3/2}}{1 \mbox{keV}} \right)^{-1} \mbox{sec},
\end{eqnarray}
which is larger than the present age of the Universe. 

The radiative decay of such a long-lived DM would contribute the
diffuse extragalactic photon background \cite{Ressel-Turner}.  The photon
number flux emitted by the gravitino is estimated as
\begin{eqnarray}
   I_E \simeq 0.4 \times 10^{28} 
    \left( \frac{E_{\gamma}}{m_{3/2}/2} \right)^{3/2}
    \mbox{cm}^{-2} \mbox{sec}^{-1} \mbox{str}^{-1}
    \left( \frac{1 \mbox{sec}}{\tau} \right)
    \left( \frac{1 \mbox{keV}}{m_{3/2}} \right)
\end{eqnarray}
for $h=0.5$ and we have assumed that the total gravitino mass density saturates
the critical density of the Universe.  Observations of the diffuse photon
background put a constraint on the lifetime
\begin{eqnarray}
    \tau \magcir 10^{25}-10^{26} \mbox{sec}
\label{lifetime-grav}
\end{eqnarray}
for $m_{3/2} \simeq 0.3-1$keV, implying $\alpha_{\lambda} \mincir
10^{-5}$.   Therefore the gravitino with lifetime longer than
Eq.~(\ref{lifetime-grav}) will be accounted for a viable DM.

The abundance of the non-thermal population of the gravitino may be affected by
the $R$ parity violation.  A necessary condition to avoid this failure is that
the NSP (the bino in our case) dominantly decays to the gravitino, not to $R$
even particles.  To illustrate how this condition puts constraints on $R$
parity breaking interactions, consider again the term $\lambda_{ijk} L_iL_j
E^c_k$ in the superpotential.  Through this interaction, the bino can decay to
three leptons, one (anti-)neutrino and two charged leptons.  We can estimate
this $R$ violating decay rate as
\begin{eqnarray}
    \frac{\alpha \alpha_{\lambda}}{192 \pi \cos^2 \theta_W}
  \times \frac{m_{\tilde b}^5}{m_{\tilde l}^4}.  \label{bino-width-R}
\end{eqnarray}
We then require that Eq.~(\ref{bino-width-R}) should be smaller than the
partial decay width to a gravitino and a photon given by
Eq.~(\ref{bino-width}), yielding a constraint
\begin{eqnarray} 
    \alpha_{\lambda} < 4 \cos^4 \theta_W \frac{m_{\tilde l}^4}{m_{3/2}^2 M^2},
\end{eqnarray}
which is $O(10^{-7}$) for the parameter range we are considering.

To conclude, we have considered the case where the light gravitinos frozen out
from the thermal bath constitute a warm DM, whereas the non-thermal gravitinos
produced by the NSP decay contribute as a volatile component of the DM.  This
mixed DM scenario requires a certain pattern of superparticle mass spectrum,
which may be tested in future collider experiments.  From the 
astrophysical side, the main implications of our DM scenario can
be summarized in the two following points: {\em (a)} as for the
primordial nucleosynthesis, the effective extra neutrino generations
associated to gravitinos imply a rather large abundance of $^4$He and D;
{\em (b)} as for large-scale structure formation, the residual
free-streaming of the ``warm" thermal gravitinos delay the galaxy
formation epoch with respect to the ``cold" case, so that the observed
high-redshift structures are required to be  associated to dwarf
protogalaxies. 

\section*{Acknowledgment}
One of the authors (MY) is grateful to S.~Davidson and G.~Raffelt for helpful
discussions, especially on constraints on the R parity violation. AM 
thanks G. Giudice for useful comments on models with low-energy SUSY 
breaking. 
The work of MY was supported in part by the Sonderforschungsbereich 375-95:
``Research in Particle-Astrophysics" of the Deutsche Forschungsgemeinschaft 
and the ECC under contracts No. SC1-CT91-0729 and No. SC1-CT92-0789.

\end{document}